\documentclass[twocolumn]{article}
\usepackage[margin=0.75in]{geometry}

\usepackage[english]{babel}
\usepackage[utf8]{inputenc}
\usepackage{amsmath}
\usepackage{graphicx}
\usepackage{subcaption}
\usepackage[numbered]{bookmark}
\usepackage{algorithm}
\usepackage{tikz}
\usepackage{pgfplots}
\usepackage{bm}
\usepackage{dsfont}
\usepackage[normalem]{ulem}
\usepackage[noblocks]{authblk}
\usepackage{cite}

\usetikzlibrary{calc}
\usetikzlibrary{pgfplots.fillbetween, backgrounds}
\usetikzlibrary{positioning}
\usetikzlibrary{shapes.geometric}
\usetikzlibrary{arrows.meta}
\usetikzlibrary{pgfplots.groupplots}
\usetikzlibrary{plotmarks}
\usetikzlibrary{calc}

\usepgfplotslibrary{groupplots}
\pgfplotsset{compat=newest} 

\DeclareGraphicsExtensions{.eps,.pdf,.png,.tikz}
\graphicspath{{figs/}}
\definecolor{blues1}{RGB}{198, 219, 239}
\definecolor{blues2}{RGB}{158, 202, 225}
\definecolor{blues3}{RGB}{107, 174, 214}
\definecolor{blues4}{RGB}{49, 130, 189}
\definecolor{blues5}{RGB}{8, 81, 156}

\definecolor{reds5}{RGB}{179,0,0}
\definecolor{reds4}{RGB}{227,74,51}

\definecolor{red2}{RGB}{228,26,28}
\definecolor{blue2}{RGB}{55,126,184}
\definecolor{green2}{RGB}{77,175,74}
\definecolor{purple2}{RGB}{152,78,163}
\definecolor{orange2}{RGB}{255,127,0}
\definecolor{pink2}{RGB}{247,129,191}

\pgfplotscreateplotcyclelist{colorbrewer-blues}{
	{blues2},
	{blues3},
	{blues4},
	{blues5},
}

\def\TRACK{1}

\ifdefined\TRACK
	\newcommand{\del}[1]{{\color{red}\sout{#1}}}

\else
	\newcommand{\del}[1]{}

\fi

\begin{document}
\title{Importance of Amplifier Physics in Maximizing the Capacity of Submarine Links}

\author{Jose Krause Perin}
\author{Joseph M. Kahn}
\affil{E. L. Ginzton Laboratory, Stanford University, Stanford, CA, 94305 USA}

\author{John D. Downie}
\author{Jason Hurley}
\author{Kevin Bennett}
\affil{Corning, Sullivan Park, SP-AR-02-1, Corning, NY, 14870 USA}


\twocolumn[
\begin{@twocolumnfalse}
	\maketitle
	\begin{abstract}
		\noindent\textbf{The throughput of submarine transport cables is approaching fundamental limits imposed by amplifier noise and Kerr nonlinearity. Energy constraints in ultra-long submarine links exacerbate this problem, as the throughput per fiber is further limited by the electrical power available to the undersea optical amplifiers. Recent works have studied how employing more spatial dimensions can mitigate these limitations. In this paper, we address the fundamental question of how to optimally use each spatial dimension. Specifically, we discuss how to optimize the channel power allocation in order to maximize the information-theoretic capacity under an electrical power constraint. Our formulation accounts for amplifier physics, Kerr nonlinearity, and power feed constraints. Whereas recent works assume the optical amplifiers operate in deep saturation, where power-conversion efficiency (PCE) is high, we show that given a power constraint, operating in a less saturated regime, where PCE is lower, supports a wider bandwidth and a larger number of spatial dimensions, thereby maximizing capacity. This design strategy increases the capacity of submarine links by about 70\% compared to the theoretical capacity of a recently proposed high-capacity system.} 
	\end{abstract}
	\vspace{0.25cm}
\end{@twocolumnfalse}
]
	
\maketitle

\section{Introduction}
Submarine transport cables interconnect countries and continents, forming the backbone of the Internet. Over the past three decades, pivotal technologies such as erbium-doped fiber amplifiers (EDFAs), wavelength-division multiplexing (WDM), and coherent detection employing digital compensation of fiber impairments have enabled the throughput per cable to jump from a few gigabits per second to tens of terabits per second, fueling the explosive growth of the information age. 

Scaling the throughput of submarine links is a challenging technical problem that has repeatedly demanded innovative and exceptional solutions. This intense technical effort has exploited a recurring strategy: to force ever-larger amounts of information over a small number of single-mode fibers \cite{Kahn2017}. This strategy is reaching its limits, however, as the amount of information that can be practically transmitted per fiber approaches fundamental limits imposed by amplifier noise and Kerr nonlinearity \cite{Mitra2001, Essiambre2010}. In submarine cables longer than about 5,000 km, this strategy faces another fundamental limit imposed by energy constraints, as the electrical power available to the undersea amplifiers ultimately restricts the optical power and throughput per fiber. 

Insights from Shannon's capacity offers a different strategy. Capacity scales linearly with the number of dimensions and only logarithmically with the power per dimension, so a power-limited system should employ more spatial dimensions (fibers or modes), while transmitting less data in each. This principle was understood as early as 1973 (see \cite{RevModPhys.66.481} and references therein) and recently embraced by the optical communications community \cite{Winzer2011, Essiambre2012, Li2014}. In fact, numerous recent works have studied how this new strategy improves the capacity and power efficiency of ultra-long submarine links \cite{Desbruslais2015, Sinkin2018, Domingues2017, Dar2017}.  But a fundamental question remained unanswered: what is the optimal way of utilizing each spatial dimension? Formally, what is the channel power allocation that maximizes the information-theoretic capacity per spatial dimension given a constraint in the total electrical power? In this paper, we formulate this problem mathematically and demonstrate how to solve it. 

In contrast to the existing literature \cite{Desbruslais2015, Domingues2017, Sinkin2018, Dar2017}, we model the optical amplifier using amplifier rate equations rather than models assuming a constant power-conversion efficiency (PCE). We argue that modeling amplifier physics is critical for translating energy constraints into parameters that govern the system capacity, such as amplification bandwidth, noise, and optical power.  This is particularly critical when the number of spatial dimensions is large, and the amplifier must operate with reduced pump power. Under these unusual operational conditions, simple constant-PCE models may not be accurate. 

Our formulation results in a non-convex optimization problem, but its solutions are robust, i.e., they do not seem to depend on initial conditions. This suggests that the optimization reaches the global minimum or is consistently trapped in an inescapable local minimum. In either case, the solutions are very promising. The optimized power allocation increases the theoretical capacity per fiber by $70\%$ compared to recently published results that employ spatial-division multiplexing (SDM) and flat power allocation. 


Our optimization yields insights into power-limited submarine link design and operation. In agreement with prior work \cite{Dar2017}, we find that overall cable capacity is maximized by employing tens of spatial dimensions per direction. Prior work, however, modeled EDFAs using a constant PCE value consistent with operation in a highly saturated regime, which maximizes PCE. Our work shows that operation in a less saturated regime, where PCE is lower, increases the useful amplification bandwidth, i.e., the number of wavelength channels for which the gain exceeds the span attenuation, and makes more power available for additional spatial dimensions. Thus, our optimization increases capacity by better utilization of both wavelength and spatial domains. Moreover, our optimization, by including Kerr nonlinearity and not neglecting it \textit{a priori}, clarifies the conditions under which nonlinearity is important, and is applicable to systems in which the number of spatial dimensions is constrained.


The remainder of this paper is organized as follows. In Section II we formulate the optimization problem and describe how to solve it. In Section III we present simulation results comparing the optimized channel power profile with conventional flat allocation designs. We conclude the paper in Section IV. We provide an Appendix on modeling EDFA physics and Kerr nonlinearity, and on the optimization algorithm.

\section{Problem formulation} \label{sec:problem}

\begin{figure}[!t]
	\flushleft
	\includegraphics[width=0.5\textwidth]{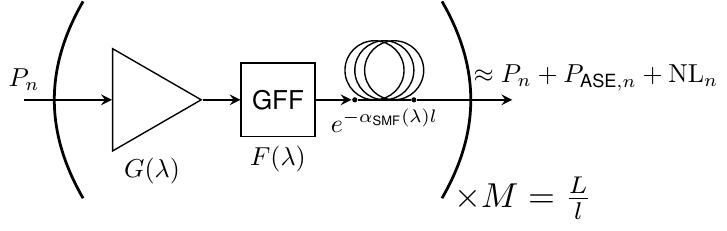}
	\caption{Equivalent block diagram of each spatial dimension of a submarine optical link including amplifier noise and nonlinear noise.} \label{fig:eq-diagram}	
\end{figure}

A submarine transport cable employs $S$ spatial dimensions in each direction, which could be modes in a multimode fiber, cores of a multi-core fiber, or simply multiple single-mode fibers. Throughout this paper, we assume that each spatial dimension is a single-mode fiber, since this is the prevailing scenario in today's submarine systems. Each of those fibers can be represented by the equivalent diagram shown in Fig.~\ref{fig:eq-diagram}.

The link has a total length $L$, and it is divided into $M$ spans, each of length $l = L/M$. An optical amplifier with gain $G(\lambda)$ compensates for the fiber attenuation $A(\lambda) = e^{\alpha_{\text{SMF}}(\lambda)l}$ of each span, and a gain-flattening filter (GFF) with transfer function $0 < F(\lambda) < 1$ ensures that the amplifier gain matches the span attenuation, so that at each span we have $G(\lambda)F(\lambda)A^{-1}(\lambda) \approx 1$. In practice, this condition has to be satisfied almost perfectly, as a mismatch of just a tenth of a dB would accumulate to tens of dBs after a chain of hundreds of amplifiers. As a result, in addition to GFF per span, periodic power rebalancing after every five or six spans corrects for any residual mismatches. 

The input signal consists of $N$ potential WDM channels spaced in frequency by $\Delta f$, so that the channel at wavelength $\lambda_n$ has power $P_n$. Our goal is to find the power allocation $P_1, \ldots, P_N$ that maximizes the information-theoretic capacity per spatial dimension. We do not make any prior assumptions about the amplifier bandwidth, hence the optimization may result in some channels not being used i.e., $P_n = 0$ for some $n$. 

Due to GFFs and periodic power rebalancing, the output signal power of each channel remains approximately constant along the link. But the signal at each WDM channel is corrupted by amplifier noise $P_{\text{ASE}, n}$ and nonlinear noise $\mathrm{NL}_n$. Thus, the $\mathrm{SNR}_n$ of the $n$th channel is given by 
\begin{equation} \label{eq:SNR}
\mathrm{SNR}_n = \begin{cases}
\displaystyle\frac{P_n}{P_{\text{ASE}, n} + \mathrm{NL}_n}, & G(\lambda_n) > A(\lambda_n) \\
0, & \text{otherwise}
\end{cases}.
\end{equation}
Note that only channels for which the amplifier gain is greater than the span attenuation can be used to transmit information, i.e., $P_n \neq 0$ only if $G(\lambda_n) > A(\lambda_n)$. 

The optical amplifiers for submarine links generally consist of single-stage EDFAs with redundant forward-propagating pump lasers operating near 980 nm. In ultra-long links, the pump power is limited by feed voltage constraints at the shores. From the maximum power transfer theorem, the total electrical power available to all undersea amplifiers is at most $\mathrm{P} = \mathrm{V}^2/(4L\rho)$, where $\mathrm{V}$ is the feed voltage, and $\rho$ is the cable resistance. To translate this constraint on the total electrical power into a constraint on the optical pump power $P_p$ per amplifier, we use an affine model similar to the one used in \cite{Desbruslais2015, Domingues2017}:
\begin{equation} \label{eq:pump-power}
P_p = \eta\Big(\frac{\mathrm{P}}{2SM} - \mathrm{P_o}\Big),
\end{equation}
where $\eta$ is an efficiency constant that translates electrical power into optical pump power, and $\mathrm{P_o}$ is a power overhead term that accounts for electrical power spent in operations not directly related to optical amplification such as pump laser lasing threshold, monitoring, and control. The factor of $2S$ appears because there are $S$ spatial dimensions in each direction.

This constraint on the pump power limits the EDFA output optical power and bandwidth, thus imposing a hard constraint on the fiber throughput. As an example, increasing $P_n$ may improve the SNR and spectral efficiency of some WDM channels, but increasing $P_n$ also depletes the EDF and reduces the amplifier overall gain. As a result, the gain of some channels may drop below the span attenuation, thus reducing the amplifier bandwidth and the number of WDM channels that can be transmitted. Further increasing $P_n$ may reduce the SNR, as the nonlinear noise power becomes significant. These considerations illustrate how forcing more power per fiber is an ineffective strategy in improving the capacity per fiber of power-limited submarine cables.

To compute the amplifier noise $P_{\text{ASE}, n}$ in a bandwidth $\Delta f$ after a chain of $M$ amplifiers, we use the analytical noise model discussed in Appendix~\ref{app:edfa}:
\begin{equation} \label{eq:amp-noise}
P_{\text{ASE}, n} = M\mathrm{NF}_nh\nu_n\Delta f,
\end{equation}
where $h$ is Planck's constant, $\nu_n$ is the channel frequency, and $\mathrm{NF}_n$ is the amplifier noise figure at wavelength $\lambda_n$. For amplifiers pumped at 980 nm, the noise figure is approximately gain-and-wavelength independent, and it can be computed from theory or measured experimentally. Although we focus on end-pumped single-mode EDFAs, similar models exist for multicore EDFAs \cite{Abedin2014}. Note that the accumulated ASE power in \eqref{eq:amp-noise} does not depend on the amplifier gain, as in Fig.~\ref{fig:eq-diagram} we defined $P_n$ as the input power to the amplifier, as opposed to the launched power. This convention conveniently makes the accumulated ASE independent of power gain.

To compute the amplifier gain, we use the semi-analytical model given in Appendix~\ref{app:edfa}. In this calculation, we assume that the input power to the amplifier is equal to $P_n + (M-1)\mathrm{NF}_nh\nu_n\Delta f$. That is, the signal power plus the accumulated ASE noise power at the input of the last amplifier in the chain. As a result, all amplifiers are designed to operate under the same conditions as the last amplifier. This pessimistic assumption is not critical in systems that operate with high optical signal-to-noise ratio (OSNR), and accounts for signal droop in low-OSNR systems, where the accumulated ASE power may be larger than the signal power, and thus reduce the amplifier useful bandwidth.

To account for Kerr nonlinearity, we use the Gaussian noise (GN) model, which establishes that the Kerr nonlinearity in dispersion-uncompensated fiber systems is well modeled as an additive zero-mean Gaussian noise whose power at the $n$th channel is given by \cite{Roberts2016}
\begin{align} \nonumber
\mathrm{NL}_n = &A^{-1}(\lambda_n)\sum_{n_1 = 1}^{N}\sum_{n_2 = 1}^{N}\sum_{q=-1}^{1} \\ \label{eq:nonlinear-noise}
& \tilde{P}_{n_1}\tilde{P}_{n_2}\tilde{P}_{n_1+n_2-n+q}D_q^{(M~\text{spans})}(n_1, n_2, n),
\end{align}
for $1 \leq n_1+n_2-n+q \leq N$. Here, $\tilde{P}_{n}$ denotes the launched power of the $n$th channel, which is related to the input power to the amplifier by $\tilde{P}_{n} = A(\lambda_n)P_n$. The nonlinear noise power is scaled by the span attenuation $A^{-1}(\lambda_n)$ due to the convention in Fig.~\ref{fig:eq-diagram} that $P_n$ refers to the input power to the amplifier, rather than the launched power. $D_q^{(M~\text{spans})}(n_1, n_2, n)$ is the set of fiber-specific nonlinear coefficients that determine the strength of the four-wave mixing component that falls on channel $n$, generated by channels $n_1$, $n_2$, and $n_1 + n_2 -n + q$. Here, $q =0$ describes the dominant nonlinear terms, while the coefficients $q = \pm 1$ describe corner contributions. These coefficients were computed in \cite{Roberts2016} and are detailed in Appendix~\ref{app:nonlinear}. 

As discussed in Section~\ref{sec:results}, for systems that operate with pump power below about 100 mW, Kerr nonlinearity is negligible and may be disregarded from the modeling.

We do not include stimulated Raman scattering (SRS) in our modeling for two reasons. First, long-haul submarine cables employ large-effective-area fibers, which reduces SRS intensity. Second, the optimized amplifier bandwidth is not larger than $45$ nm, while the Raman efficiency peaks when the wavelength difference is $\sim 100$ nm. 

Using equations \eqref{eq:SNR}--\eqref{eq:nonlinear-noise}, we can compute Shannon's capacity per fiber by adding the capacities of the individual WDM channels:
\begin{align} \label{eq:capacity}
C = 2\Delta f\sum_{n=1}^N &\mathds{1}\{\mathcal{G}(\lambda_n) \geq \mathcal{A}(\lambda_n)\}\log_2(1 + \Gamma\mathrm{SNR}_n),
\end{align}
where $0 < \Gamma < 1$ is the coding gap to capacity and $\mathcal{G}(\lambda_n), \mathcal{A}(\lambda_n)$ denote, respectively, the amplifier gain and span attenuation in dB units. The indicator function $\mathds{1}\{\cdot\}$ is one when the condition in its argument is true, and zero otherwise. As we do not know \textit{a priori} which channels contribute to capacity ($P_n \neq 0$), we sum over all channels and let the indicator function indicate which channels have gain above the span attenuation. 

Since the indicator function is non-differentiable, it is convenient to approximate it by a differentiable sigmoid function such as 
\begin{equation}
\mathds{1}\{x \geq 0\} \approx 0.5(\tanh(Dx) + 1),
\end{equation}
where $D > 0$ controls the sharpness of the sigmoid approximation. Although making $D$ large better approximates the indicator function, it results in vanishing gradients, which retards the optimization process. 

Hence, the optimization problem of maximizing the capacity per fiber given an energy constraint that limits the amplifier pump power $P_p$ can be stated as
\begin{align} \label{eq:optimization} \nonumber
\underset{L_\text{EDF}, \mathcal{P}_1, \ldots, \mathcal{P}_N}{\mathrm{maximize}}~&C \\
\text{given}~&P_p
\end{align}
In addition to the power allocation $\mathcal{P}_1, \ldots, \mathcal{P}_N$ in dBm units, we optimize over the EDF length $L_{\text{EDF}}$, resulting in a $(N+1)$-dimensional non-convex optimization problem. $L_{\text{EDF}}$ may be removed from the optimization if its value is predefined. It is convenient to optimize over the signal power in dBm units, as the logarithmic scale enhances the range of signal power that can be covered by taking small adaptation steps. Even if we assumed binary power allocation, i.e., $\mathcal{P}_n \in \{0, \bar{\mathcal{P}}\}$, it is not easy to determine the value of $\bar{\mathcal{P}}$ that will maximize the amplification bandwidth for which the gain is larger than the span attenuation.

Note that if we did not have the pump power constraint and the amplifier gain did not change with the power allocation $\mathcal{P}_1, \ldots, \mathcal{P}_N$, the optimization problem in \eqref{eq:optimization} would reduce to the convex problem solved in \cite{Roberts2016}. Therefore, we can argue that to within a small $\Delta \mathcal{P}_n$ that does not change the conditions in the argument of the indicator function, the objective \eqref{eq:capacity} is locally concave.

\begin{table}[t]
	\centering
	\caption{Simulation parameters.} \label{tab:param}
	\resizebox{0.5\textwidth}{!}{
		\begin{tabular}{l|c|c}
			\hline
			Parameter & Value & Units \\
			\hline
			Link length ($L$) & 14,350 & km \\
			Span length ($l$) & 50 & km \\
			Number of amplifiers per fiber ($M$) & 287 &  \\
			\hline
			First channel ($\lambda_1$) & 1522 & nm \\
			Last channel ($\lambda_N$) & 1582 & nm \\
			Channel spacing ($\Delta f$) & 50 & GHz \\
			Max. number of WDM channels ($N$) & 150 & \\
			\hline
			Fiber attenuation coefficient ($\alpha_\text{SMF}(\lambda)$) & 0.165 & dB km$^{-1}$ \\
			Fiber dispersion coefficient ($D(\lambda)$) & 20 & ps nm$^{-1}$ km$^{-1}$ \\
			Fiber nonlinear coefficient ($\gamma$) & 0.8 & W$^{-1}$ km$^{-1}$ \\
			Fiber additional loss (margin) & 1.5 & dB \\
			Overall span attenuation ($A(\lambda)$) & $8.25 + 1.5 = 9.75$ & dB \\
			Nonlinear noise power scaling ($\epsilon$) & 0.07 & \\
			\hline
			Coding gap ($\Gamma$) & $-1$ & dB \\
			Sigmoid sharpness ($D$) & 2 & \\
			Excess noise factor ($n_{sp}$) & 1.4 & \\
			Excess loss ($l_k$) & 0 & dB/m \\
			\hline
		\end{tabular}
	}
\end{table}

\begin{figure*}[t!]
	\centering
	\includegraphics[width=\textwidth]{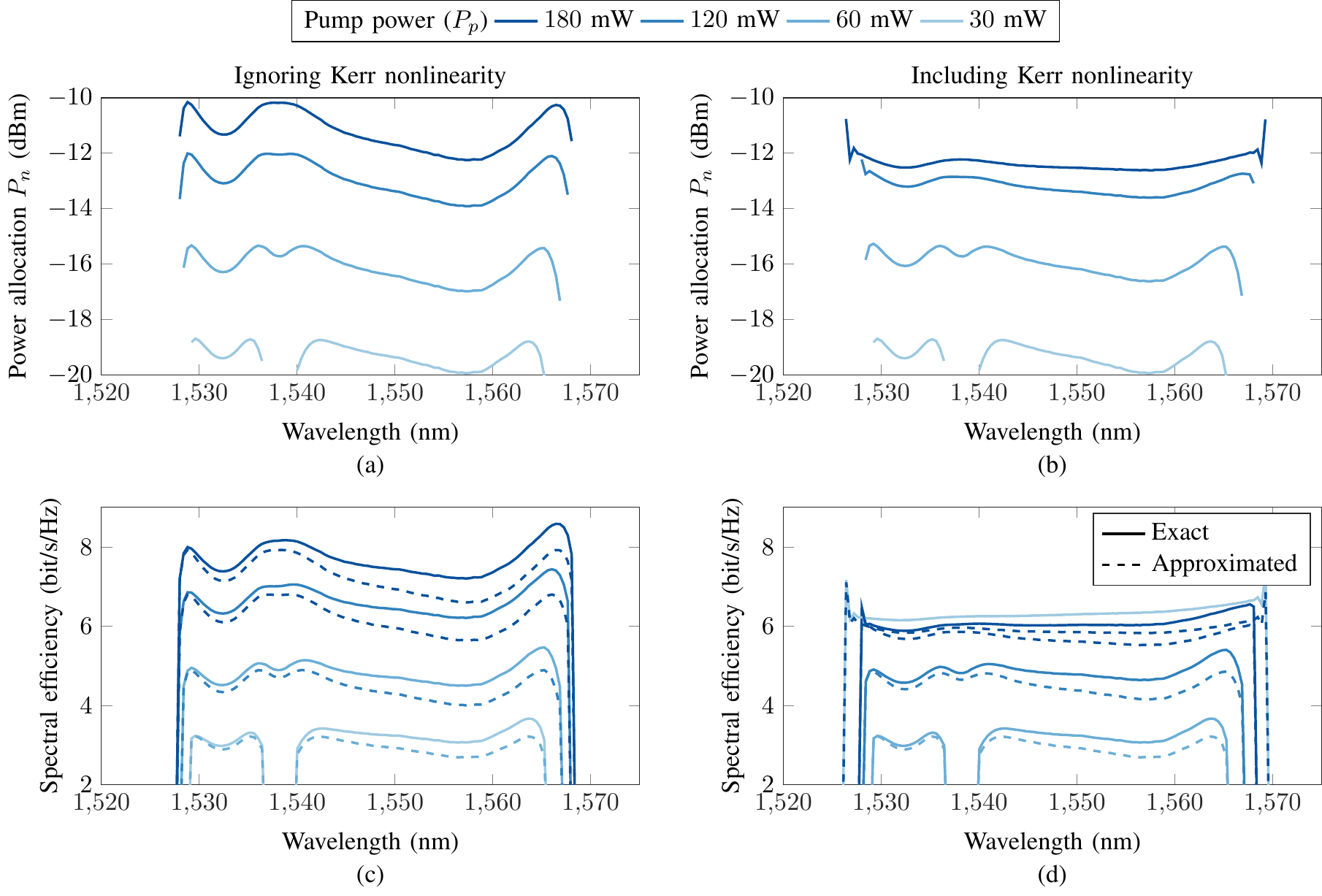}
	\caption{Optimized power allocation $P_n$ for several values of pump power $P_p$. Kerr nonlinearity is disregarded in (a) and included in (b). Their corresponding achievable spectral efficiency is shown in (c) and (d). Note that $P_n$ corresponds to the input power to the amplifier. The launched power is $\tilde{P}_{n} = G(\lambda_n)F(\lambda_n)P_n = A(\lambda_n)P_n$. Thus, the launch power is 9.75 dB above the values shown in these graphs.} \label{fig:optimal_power}
\end{figure*}

Nevertheless the optimization problem in \eqref{eq:optimization} is not convex, and therefore we must employ global optimization techniques. In this paper, we use the particle swarm optimization (PSO) algorithm \cite{Kennedy1942}. The PSO randomly initializes $R$ particles $X = [L_{\text{EDF}}, \mathcal{P}_1, \ldots, \mathcal{P}_N]^T$. As the optimization progresses, the direction and velocity of each particle is influenced by its best known position and also by the best known position found by other particles in the swarm. The PSO algorithm was shown to outperform other global optimization algorithms such as the genetic algorithm in a broad class of problems \cite{hassan2005}. Further details of the PSO are given in Appendix~\ref{app:opt}. 

When nonlinear noise is negligible, the solutions found through PSO are robust. That is, they do not depend on the initial conditions. When nonlinear noise is significant, different particle initializations lead to the same overall solution, but these solutions differ by small random variations. To overcome this problem, once the PSO algorithm stops, we continue the optimization using the saddle-free Newton's (SFN) method \cite{Dauphin2014}. This variant of Newton's method is suited to non-convex problems, as it is not attracted to saddle points. It requires knowledge of the Hessian matrix, which can be computed analytically or through finite differences of the gradient. Further details of the SFN method are given in Appendix~\ref{app:opt}. 

\section{Results and Discussion} \label{sec:results}

We now apply our proposed optimization procedure to the reference system with parameters listed in Table~\ref{tab:param}. These parameters are consistent with recently published experimental demonstration of high-capacity systems employing SDM \cite{Sinkin2018}. We consider $M = 287$ spans of $l = 50$ km of low-loss large-effective area single-mode fiber, resulting in a total link length of $L = 14,350$ km. The span attenuation is $\mathcal{A}(\lambda) = 9.75$ dB, where 8.25 dB is due to fiber loss, and the additional 1.5 dB is added as margin. For the capacity calculations we assume a coding gap of $\Gamma = 0.79$ ($-1$ dB).

\subsection{Optimized channel power}
We first study how the optimized power allocation and the resulting spectral efficiency is affected by the amplifier pump power. We also investigate how Kerr nonlinearity affects the optimized power allocation and when it can be neglected. This discussion does not assume any particular electrical power budget or number of spatial dimensions. In Section~\ref{sec:optimal-dim}, we consider how employing multiple spatial dimensions can lead to higher overall cable capacity.

For a given pump power $P_p$, we solve the optimization problem in \eqref{eq:optimization} for the system parameters listed in Table~\ref{tab:param}. The resulting power allocation $P_n$ is plotted in Fig.~\ref{fig:optimal_power} when Kerr nonlinearity is (a) disregarded and (b) included. The corresponding achievable spectral efficiency of each WDM channel is shown in Fig.~\ref{fig:optimal_power}cd. 

\begin{figure*}[t]
	\centering
	\includegraphics[width=\textwidth]{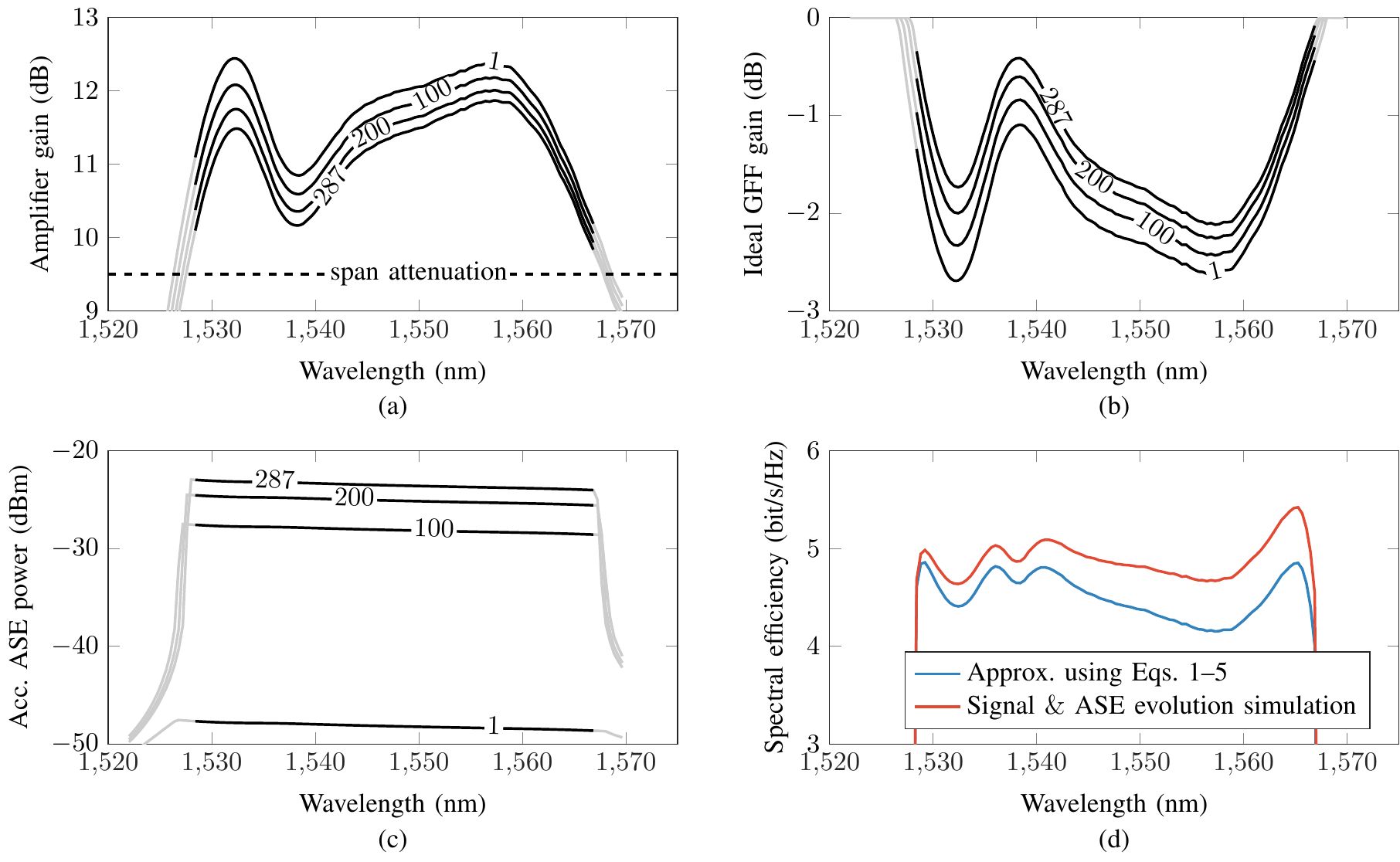}
	\caption{Theoretical (a) amplifier gain, (b) ideal GFF gain, and (c) accumulated ASE power in 50 GHz after 1, 100, 200, and 287 spans of 50 km. The pump power of each amplifier is 60 mW, resulting in the optimized power profile shown in Fig.~\ref{fig:optimal_power}b for $P_p = 60$ mW and EDF length of 6.27 m.} \label{fig:ase-acc-sim}
\end{figure*}


For small pump powers, the optimized power profile is limited by amplifier properties, and thus there is only a very small difference between the two scenarios shown in Fig.~\ref{fig:optimal_power}. As the pump power increases and the amplifier delivers more output power, Kerr nonlinearity becomes a factor limiting the channel power. Interestingly, the optimized power allocation in the nonlinear regime exhibits large variations at the extremities because the nonlinear noise is smaller at those channels. Although the optimization is performed for 150 possible WDM channels from 1522 nm to 1582 nm, Fig.~\ref{fig:optimal_power}a and (b) show that not all of these WDM channels can be utilized. The useful bandwidth is restricted between roughly 1528 nm and 1565 nm. Note that, as expected, the useful bandwidth does not increase significantly even when the pump power $P_p$ is tripled from 60 mW to 180 mW, since the amplification bandwidth is fundamentally limited by the EDF's gain and absorption coefficients. However, for very small pump powers the useful amplification bandwidth decreases, as the gain for some channels becomes insufficient to compensate for the span attenuation. For instance, note that for $P_p = 30$ mW, part of the amplifier bandwidth cannot be used, as the resulting amplifier gain is below attenuation. The optimized EDF length does not vary significantly, and it is generally in the range of 6 to 8 m.

The solid lines in Fig.~\ref{fig:optimal_power}c and (d) are obtained from \eqref{eq:capacity} by using exact models \eqref{eq:scd} for the amplifier gain and noise, while the dashed lines are computed by making approximations to allow (semi-)analytical calculation of amplifier gain \eqref{eq:semi_analytical_gain} and noise \eqref{eq:amp-noise} (see Appendix~\ref{app:edfa}), and speed up the optimization process. Although the approximations lead to fairly small errors in estimating the spectral efficiency, we emphasize that after the optimizations conclude, the exact models are used to definitively quantify the spectral efficiency and overall system capacity obtained. 

\subsection{Signal and ASE evolution}
For the optimized power profile for $P_p = 60$ mW shown in Fig.~\ref{fig:optimal_power}b, we compute the evolution of amplifier gain, accumulated ASE, and the required GFF gain along the 287 spans, as shown in Fig.~\ref{fig:ase-acc-sim}. The amplifier gain and ASE power are computed using the exact amplifier model given in \eqref{eq:scd}. The accumulated ASE power (Fig.~\ref{fig:ase-acc-sim}c) increases after every span, causing the amplifier gain (Fig.~\ref{fig:ase-acc-sim}a) and consequently the ideal GFF gain (Fig.~\ref{fig:ase-acc-sim}b) to change slightly. 

Note that the ideal GFF profiles have ripples of less than 3 dB. The variations in amplifier gain along the 287 spans are also small, resulting in GFF shape difference of less than 2 dB between the first and last GFFs. In practice, the ideal GFF shape can be achieved by fixed GFF after each amplifier and periodic power rebalancing at intervals of five or so spans. As a test of how critically important ideal per-span gain flattening is, we considered a scenario in which all GFFs are identical to the last GFF (labeled 287 in Fig.~\ref{fig:ase-acc-sim}b), and ideal power rebalancing is realized only after every 10 spans. The difference in capacity in this scenario is less than $3\%$ with respect to the ideal case. 

At the last span of the signal and ASE evolution simulation, we compute the spectral efficiency per channel and compare it to the approximated results obtained using \eqref{eq:SNR}--\eqref{eq:capacity}. As in the discussion of Fig.~\ref{fig:optimal_power}(c) and (d), although the approximations lead to fairly small errors in estimating the spectral efficiency, we use the exact calculations to definitively quantify spectral efficiency and overall capacity.


\begin{figure*}[t]
	\centering
	\def\CITATION{\cite{Sinkin2018}}
	\includegraphics[width=\textwidth]{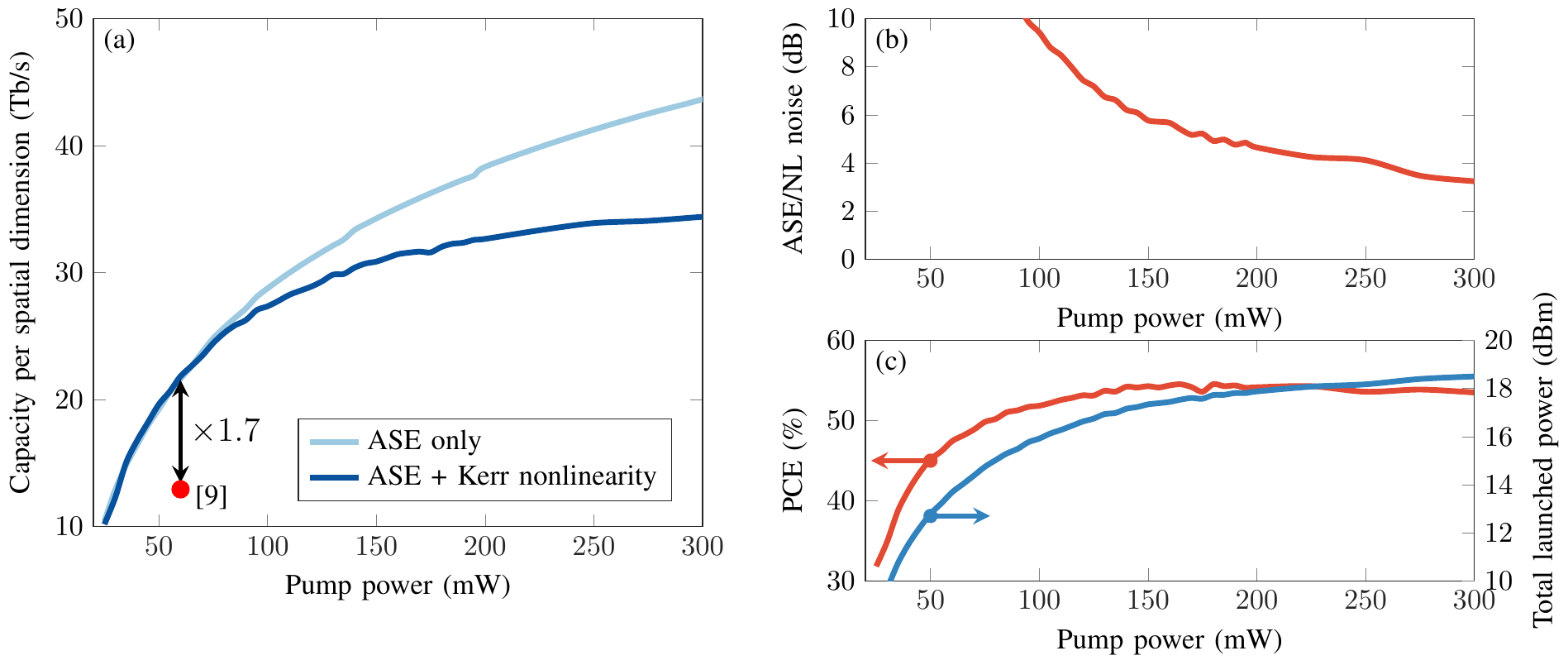}
	\caption{(a) Total capacity per single-mode fiber as a function of pump power. The power allocation and EDF length are optimized for each point. The red dot corresponds to the capacity according to \eqref{eq:capacity} for a system with parameters consistent with \cite{Sinkin2018}. (b) Corresponding ratio between ASE and nonlinear noise power, and (c) corresponding power conversion efficiency and total launched optical power.} \label{fig:capacity}
\end{figure*}

\subsection{Capacity per spatial dimension}
Fig.~\ref{fig:capacity}a shows the total capacity per fiber as a function of the pump power. Once again, for each value of pump power $P_p$, we solve the optimization problem in \eqref{eq:optimization} for the system parameters listed in Table~\ref{tab:param}. The capacity per spatial dimension plotted in Fig.~\ref{fig:capacity}a is computed by summing the capacities of the individual WDM channels. Below about 100 mW of pump power, the system operates in the linear regime. At higher pump powers, the amplifier can deliver higher optical power, but Kerr nonlinearity becomes significant and limits the capacity. Fig.~\ref{fig:capacity}b details the ratio between ASE power and nonlinear noise power. At high pump powers, ASE is only about 4 dB higher than nonlinear noise. This illustrates the diminishing returns of forcing more power over a single spatial dimension.

Fig.~\ref{fig:capacity}c shows the total launched optical power and amplifier power conversion efficiency (PCE) defined as $\mathrm{PCE}\equiv \frac{\text{total output optical power} - \text{total input optical power}}{\text{optical pump power}}$ \cite{Desurvire1995}. From energy conservation arguments, it can be shown that PCE is upper bounded by the ratio between pump and signal wavelengths, which for 980 nm pump results in $\mathrm{PCE} < 63 \%$ \cite{Desurvire1995}. Fig.~\ref{fig:capacity}c also shows the diminishing returns of forcing more power over a single spatial dimension, since the amplifier efficiency does not increase linearly with pump power. In fact, doubling the pump power from 50 mW to 100 mW increases PCE by only $7.43 \%$. Clearly, this additional pump power could be better employed in doubling the number of spatial dimensions, which would nearly double the overall cable capacity.

We have also computed the capacity for different span lengths assuming a total pump power per fiber of $P_{p, total} = 287\times 50 = 14350$ mW. In agreement to prior work \cite{Downie2018}, the optimal span length is achieved for 40--50 km, resulting in an optimal span attenuation of 8.1--9.75 dB.

\subsection{Comparison to experimental system}
To gauge the benefits of our proposed optimization procedure, we compare the results of our approach to those of a recently published work \cite{Sinkin2018}, which experimentally demonstrated high-capacity SDM systems. In their experimental setup, Sinkin et al used 82 channels spaced by 33 GHz from 1539 nm to 1561 nm. Each of the 12 cores of the multicore fiber was amplified individually by an end-pumped EDFA with forward-propagating pump. Each amplifier was pumped near 980 nm with 60 mW resulting in an output power of 12 dBm \cite{Sinkin2018}, thus $-7.1$ dBm per channel. The span attenuation was 9.7 dB, leading to the input power to the first amplifier of $P_n = -16.7$ dBm per channel.  We compute the capacity of this system according to \eqref{eq:capacity} using the same methods and models for amplifier and Kerr nonlinearity discussed in Section~\ref{sec:problem}. Fiber parameters and amplifier noise figure are given in Table~\ref{tab:param}. The EDF length is assumed 7 m, which is the value resulting from our optimization for EDFAs pumped with 60 mW. The resulting achievable spectral efficiency per channel is, on average, 4.8 bit/s/Hz, yielding a maximum rate of about 13 Tb/s per core. This is indicated by the red dot in Fig.~\ref{fig:capacity}. Naturally, this calculation is oversimplified, but it is consistent with the rate achieved in \cite{Sinkin2018}. Their experimental spectral efficiency is 3.2 bit/s/Hz in 32.6 Gbaud, leading to 106.8 Gb/s per channel, 8.2 Tb s$^{-1}$ per core, and 105 Tb s$^{-1}$ over the 12 cores. The capacity using the optimized power profile is about 22 Tb s$^{-1}$ per core for the same pump power and overall system (ASE + Kerr nonlinearity curve in Fig.~\ref{fig:capacity}), thus offering $70 \%$ higher capacity when compared to the theoretical estimate for a system consistent with \cite{Sinkin2018}. The optimized power profile for $P_p = 60$ mW is plotted in Fig.~\ref{fig:optimal_power}b.


The main benefit of the channel power optimization is to allow the system to operate over a wider amplification bandwidth with more spatial dimensions. Capacity scales linearly with the number of dimensions (frequency or spatial) and only logarithmically with power. The optimization tends to favor lower signal powers, inducing less gain saturation and allowing higher gain for a given pump power. This increases the usable bandwidth, over which the gain exceeds the span attenuation, and frees pump power for additional spatial dimensions. The optimization does not necessarily optimize the amplifiers for high PCE. Highly saturated optical amplifiers achieve higher PCE, but that does not necessarily translate to higher power-limited information capacity.

\subsection{Optimal number of spatial dimensions} \label{sec:optimal-dim}
\begin{figure}[t!]
	\centering
	\includegraphics[width=0.5\textwidth]{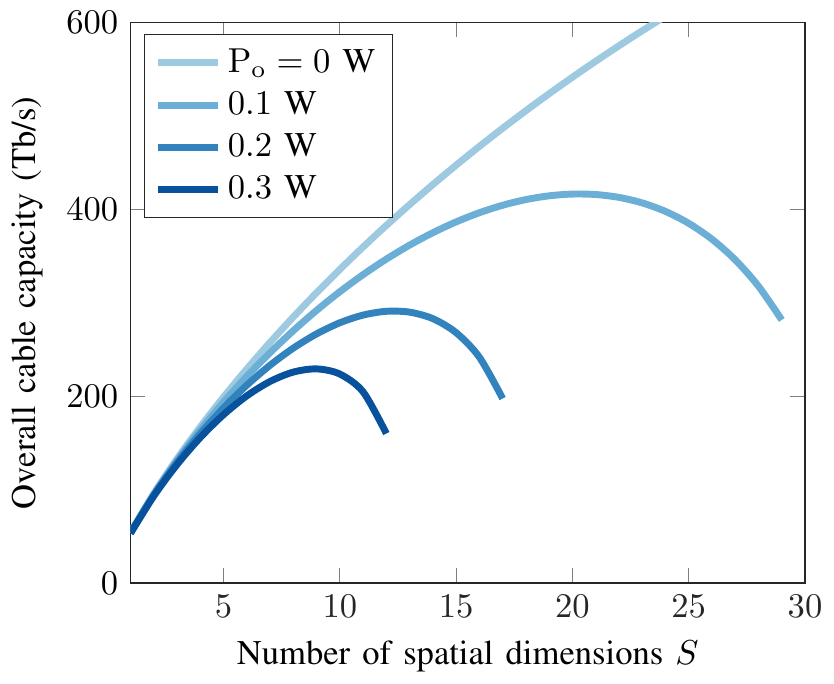}
	\caption{Capacity as a function of the number of spatial dimensions for the system of Table~\ref{tab:param} assuming a power budget of $\mathrm{P} = 2.5$ kW for all amplifiers.} \label{fig:spatial-dims}
\end{figure}

The optimal strategy is therefore to employ more spatial dimensions while transmitting less power in each one. The optimal number of spatial dimensions depends on the available electrical power budget. As an example, Fig.~\ref{fig:spatial-dims} shows the capacity of a cable employing $S$ spatial dimensions in each direction. We consider the feed voltage $\mathrm{V} = 12$ kV, cable resistivity $\rho = 1~\Omega$ km$^{-1}$, and the reference link of Table~\ref{tab:param}. Thus, the total electrical power available for all amplifiers is 2.5 kW. From this and assuming efficiency $\eta = 0.4$ and overhead power $\mathrm{P_o}$, we can compute the pump power per amplifier $P_p$ according to \eqref{eq:pump-power}, and obtain the capacity per fiber from Fig.~\ref{fig:capacity}a.

The optimal number of spatial dimensions in each direction $S$ decreases as the overhead power increases, reaching 20, 12, and 8 for the power overhead $\mathrm{P_o} = 0.1, 0.2$, and 0.3 W, respectively. This corresponds to amplifiers with pump powers of 43.7, 47.4, and 65.7 mW, respectively. Hence, at the optimal number of spatial dimensions the system operates in the linear regime, as can be seen by inspecting Fig.~\ref{fig:capacity}. For small values of $\mathrm{P_o}\to 0$, the optimal number of spatial dimensions is very large, illustrating the benefits of massive SDM, as reported in \cite{Dar2017}.

Fig.~\ref{fig:spatial-dims} also illustrates the diminishing returns of operating at a very large number of spatial dimensions. Consider, for instance, the curve for power overhead $\mathrm{P_o} = 0.1$ W. The optimal number of spatial dimensions is $S = 20$, resulting in a total capacity per cable of about 383 Tb s$^{-1}$. However, with half of this number of spatial dimensions $S = 10$ (and $P_p = 135$ mW), we can achieve about $80 \%$ of that capacity. Thus, systems subject to practical constraints such as cost and size may operate with a number of spatial dimensions that is not very large.

\begin{figure}[t!]
	\centering
	\includegraphics[width=0.5\textwidth]{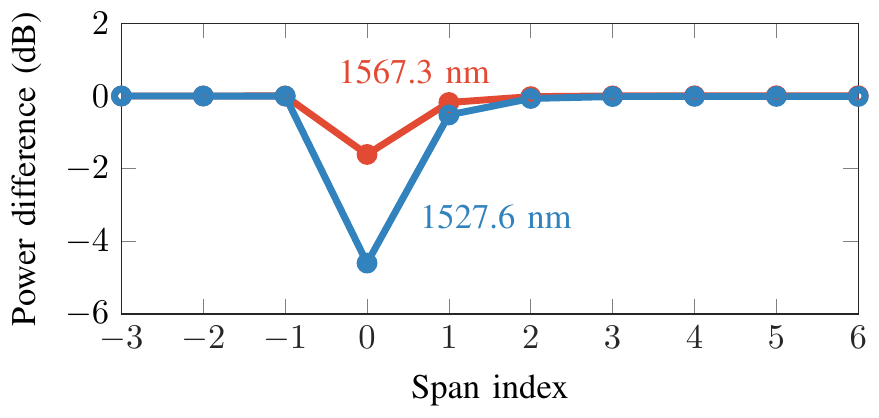}
	\caption{Difference in signal power with respect to correct power allocation in the event of a single pump failure at the span indexed by zero. After about two spans the power levels are restored to their correct values.} \label{fig:pump-fail}
\end{figure}

\subsection{Recovery from pump failure}
An important practical consideration for submarine systems is their ability to recover when the input power drops significantly due to faulty components or pump laser failure. Thus, submarine amplifiers are designed to operate in high gain compression, so that the power level can recover from these events after a few spans. We show that the optimized input power profile and amplifier operation can still recover from such events. Fig~\ref{fig:pump-fail} illustrates the power variation with respect to the optimized power profile when one of the two pump lasers in an amplification module fails. The failure occurs at the span indexed by zero. The amplifier operates with redundant pumps resulting in $P_p = 50$ mW, and in the event of a single-pump failure the power drops to $P_p = 25$ mW. The signal power in the channels at the extremities of the spectrum are restored with just two spans. Capacity is not significantly affected by a single-pump failure, since the amplifier noise increases by less than 0.5 dB in all channels. Although the power levels could still be restored in the event that the two pump lasers in the module fail, the total amplifier noise power would be about 10 dB higher in some channels.

\section{Conclusion}

We have demonstrated how to maximize the information-theoretic capacity of ultra-long submarine systems by optimizing the channel power allocation in each spatial dimension. Our models account for EDFA physics, Kerr nonlinearity, and power feed limitations. Modeling EDFA physics is paramount to understanding the effects of energy limitations on amplification bandwidth, noise, and optical power, which intimately govern the system capacity. We show that this optimization results in 70\% higher capacity when compared to the theoretical capacity of a recently proposed high-capacity system. Our optimization also provides insights on the optimal number of spatial dimensions, optimal amplifier operation, and the impact of Kerr nonlinearity. Our proposed technique could be used in optimizing existing systems, and also to design future systems leveraging SDM.

\appendix
\section{Amplifier physics} \label{app:edfa}

\begin{figure}[t!]
	\centering
	\includegraphics[width=0.5\textwidth]{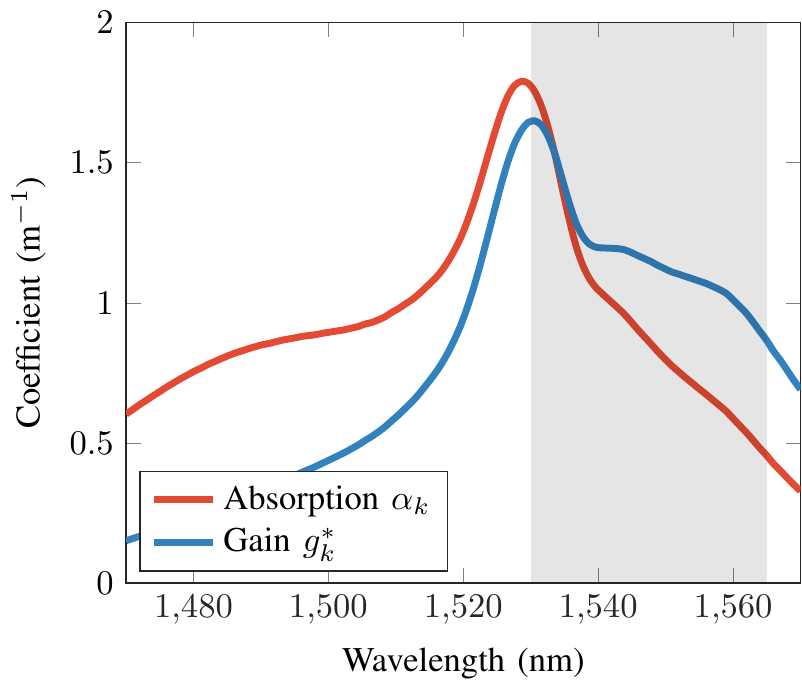}
	\caption{Absorption and gain coefficients for the EDF used in this paper. C band is highlighted. For the pump at 980 nm, $\alpha_p = 0.96$ m$^{-1}$, and $g_p^* = 0$ m$^{-1}$. Other relevant parameters are $\mathrm{NA} = 0.28$, $r_{Er} = 0.73 \mu\mathrm{m}$, and $\bar{n}_t = 9.96\times 10^{18}$ cm$^{3}$.} \label{fig:eqf-abs-gain}
\end{figure}

The steady-state pump and signal power evolution along an EDF of length $L_{EDF}$ is well modeled by the standard confined-doping (SCD) model \cite{Giles1991}, which for a two-level system is described by a set of coupled first-order nonlinear differential equations:

\begin{align} \label{eq:scd} \nonumber
\frac{d}{dz}P_k(z) &= u_k(\alpha_k + g^*_k)\frac{\bar{n}_2}{\bar{n}_t}P_k(z) \\
&- u_k(\alpha_k+l_k)P_k(z) + 2u_kg^*_k\frac{\bar{n}_2}{\bar{n}_t}h\nu_k\Delta f
\end{align}

\begin{equation}
\frac{\bar{n}_2}{\bar{n}_t} = \frac{\sum_k \frac{P_k(z)\alpha_k}{h\nu_k\zeta}}{1 + \sum_k \frac{P_k(z)(\alpha_k + g^*_k)}{h\nu_k\zeta}}
\end{equation}
where the subindex $k$ indexes both signal and pump i.e., $k \in \{p, 1, \ldots, N\}$, $z$ is the position along the EDF, $l_k$ is the excess loss, and $u_k = 1$ for beams that move in the forward direction i.e., increasing $z$, and $u_k = -1$ otherwise. $\alpha_k$ is the absorption coefficient, $g^*_k$ is the gain coefficient, and $\bar{n}_2/\bar{n}_t$ denotes the population of the second metastable level normalized by the Er ion density $\bar{n}_t$. $\zeta = \pi r_{Er}^2\bar{n}_t/\tau$ is the saturation parameter, where $r_{Er}$ is the Er-doping radius, and $\tau \approx 10$ ms is the metastable lifetime. According to this model, the amplifier characteristics are fully described by three macroscopic parameters, namely $\alpha_k$, $g^*_k$, and $\zeta$. Fig~\ref{fig:eqf-abs-gain} shows $\alpha_k$ and $g^*_k$ for the EDF used in our simulations for this paper.
The first term of \eqref{eq:scd} corresponds to the medium gain, the second term accounts for absorption, and the third term accounts for amplified spontaneous emission (ASE) noise. 

To compute the amplifier gain and noise using \eqref{eq:scd}, we must solve the boundary value problem (BVP) of $N + 1 + 2N$ coupled equations, where we have $N$ equations for the signals, one for the pump, and the noise at the signals' wavelengths is broken into $2N$ equations: $N$ for the forward ASE, and $N$ for the backward ASE. 

Although \eqref{eq:scd} is very accurate, the optimizations require evaluation of the objective function hundreds of thousands of times, which would require solving the BVP in \eqref{eq:scd} that many times. Hence, approximations for the gain and noise are necessary. 

By assuming that the amplifier is not saturated by ASE, equation \eqref{eq:scd} reduces to a single-variable implicit equation \cite{Saleh1990}, which can be easily solved numerically. According to this model, the amplifier gain is given by 

\begin{equation} \label{eq:semi_analytical_gain}
G_k = \exp\Big(\frac{\alpha_k + g^*_k}{\zeta}(Q^{in} - Q^{out}) - \alpha_kL_{\text{EDF}}\Big)
\end{equation}
where $Q^{in}_k = \frac{P_k}{h\nu_k}$ is the photon flux in the $k$th channel, and  $Q^{in} = \sum_k Q^{in}_k$ is the total input photon flux. The output photon flux $Q^{out}$ is given by the implicit equation:
\begin{equation} \label{eq:out-photon-flux}
Q^{out} = \sum_kQ^{in}_k\exp\Big(\frac{\alpha_k + g^*_k}{\zeta}(Q^{in} - Q^{out}) - \alpha_kL_{\text{EDF}}\Big)
\end{equation}

Therefore, to compute the amplifier gain using the semi-analytical model, we must first solve \eqref{eq:out-photon-flux} numerically for $Q^{out}$, and then compute the gain using \eqref{eq:semi_analytical_gain}. This procedure is much faster than solving \eqref{eq:scd}.

The semi-analytical model is useful to compute the gain, but it does not give us any information about the noise power. Thus, we must use a further simplification. By assuming that the amplifier is inverted uniformly, equation \eqref{eq:scd} can be solved analytically resulting in the well-known expression for ASE power in a bandwidth $\Delta f$ for a single amplifier:
\begin{equation}
P_{\text{ASE}, n} = 2n_{sp, n}(G_n - 1)h\nu_n\Delta f
\end{equation}
where $n_{sp}$ is the excess noise factor \cite[equation (32)]{Giles1991}. The excess noise factor is related to the noise figure $NF_n = 2n_{sp, n}\frac{G_n-1}{G_n}$, where the commonly used high-gain approximation
$\frac{G(\lambda_n)-1}{G(\lambda_n)} \approx 1$ may be replaced by the more accurate approximation $\frac{G(\lambda_n)-1}{G(\lambda_n)} \approx 1 - e^{-\alpha_{SMF}l}$, since in submarine systems the amplifier gain is approximately equal to the span attenuation, which is on the order of 10 dB. This approximation conveniently makes the amplifier noise figure independent of the amplifier gain.

\begin{figure*}[t]
	\includegraphics[width=\textwidth]{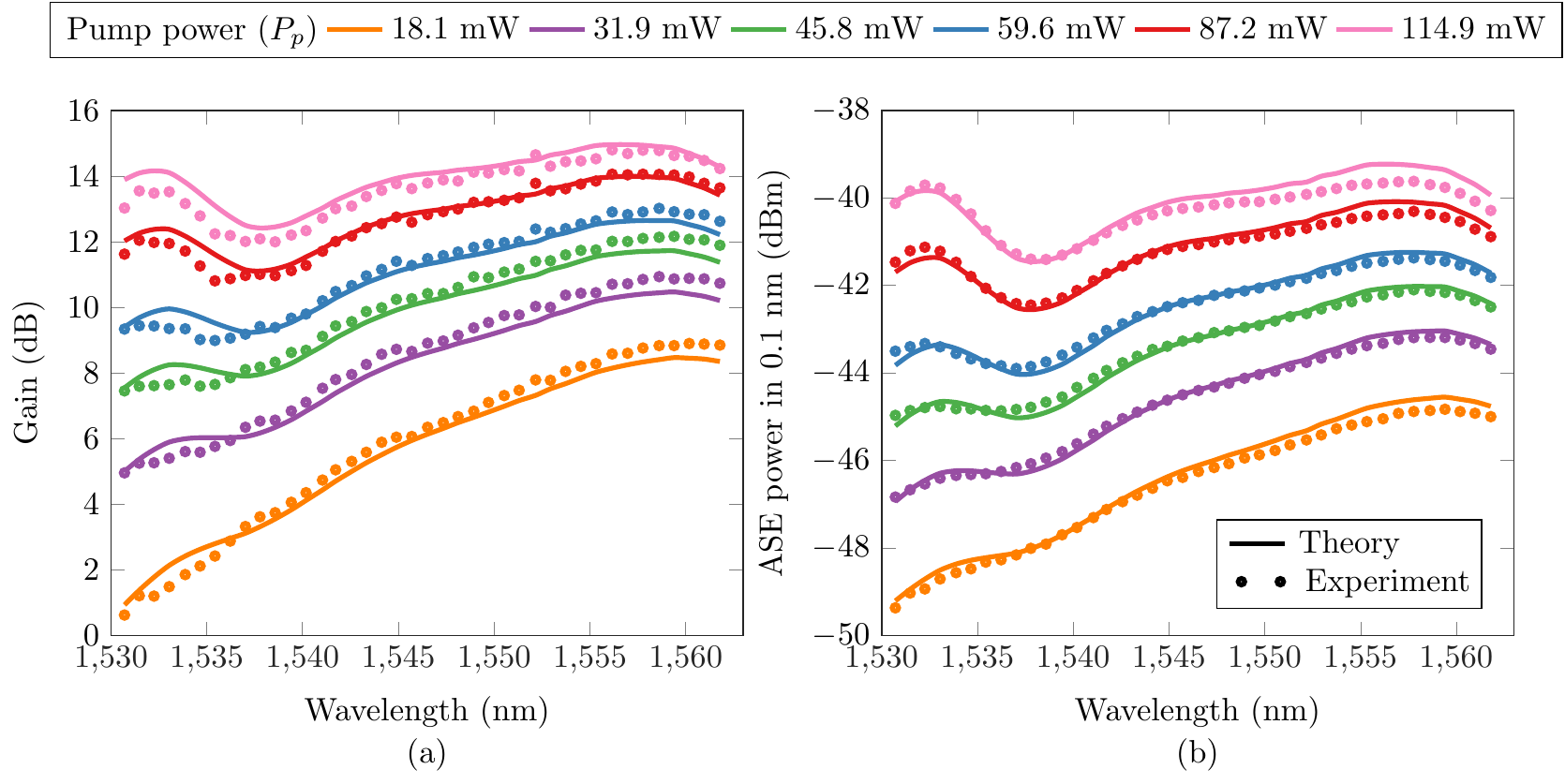}
	\caption{Comparison between experiment and theory for (a) gain and (b) ASE power in 0.1 nm for different values of pump power. Theoretical gain and ASE curves are computed using \eqref{eq:scd}. } \label{fig:experiment}
\end{figure*}

Fig.~\ref{fig:experiment} compares the gain and ASE power predicted using the theoretical model in \eqref{eq:scd} with experimental measurements for several values of pump power $P_p$. The amplifier consists of a single 8-m-long EDF pumped by a forward-propagating laser near 980 nm with power $P_p$. The incoming signal to the amplifier consists of 40 unmodulated signals from 1531 to 1562. The power of each signal is $-13$ dBm, resulting in a total of $3$ dBm. The theoretical results use \eqref{eq:scd} with experimentally measured values of the absorption and gain coefficients $\alpha$ and $g^*$. The nominal experimentally measured values have been scaled up by $8\%$ to achieve the best fit between theory and experiment. The experimental error in these values was estimated independently to be about $5\%$.

\section{Discrete Gaussian noise model} \label{app:nonlinear}

The nonlinear coefficients $D^{(1 \text{span})}_q(n_1, n_2, n)$ for one span of single-mode fiber of length $l$, nonlinear coefficient $\gamma$, power attenuation $\alpha_{\text{SMF}}$, and propagation constant $\beta_2$ are given by the triple integral

\begin{align} \nonumber 
D_q^{(1~\text{span})}(n_1, n_2, n) &= \frac{16}{27}\gamma^2\iiint_{-1/2}^{1/2} \\ \nonumber
&\rho((x + n_1)\Delta f, (y + n_2)\Delta f, (z + n)\Delta f) \\ \label{eq:Dcoeff}
&\cdot \mathrm{rect}(x+y-z+q)\partial x\partial y\partial z,
\end{align}
\begin{align}
\rho(f_1, f_2, f) &= \Bigg|\frac{1 - \exp(-\alpha l + j4\pi^2\beta_2l(f_1-f)(f_2-f))}{\alpha - j4\pi^2\beta_2(f_1-f)(f_2-f)}\Bigg|^2,
\end{align}
where $\mathrm{rect}(\omega) = 1$, for $|\omega| \leq 1/2$, and $\mathrm{rect}(\omega) = 0$ otherwise. Equation \eqref{eq:Dcoeff} assumes that all channels have a rectangular spectral shape. 

Computing $D_q^{(1~\text{span})}(n_1, n_2, n)$ is computationally less intensive than $D_q^{(M~\text{spans})}(n_1, n_2, n)$, since the highly oscillatory term $\chi(f_1, f_2, f)$ in $D_q^{(M~\text{spans})}(n_1, n_2, n)$ \cite{Roberts2016} is constant and equal to one in $D_q^{(1~\text{span})}(n_1, n_2, n)$. The nonlinear coefficients for $M$ spans can be computed by following the nonlinear power scaling given in \cite{Poggiolini2012}:

\begin{equation}
D_q^{(M~\text{spans})}(n_1, n_2, n) = M^{1 + \epsilon}D_q^{(1~\text{span})}(n_1, n_2, n),
\end{equation} 
where the parameter $\epsilon$ controls the nonlinear noise scaling over multiple spans, and for bandwidth of $\sim 40$ nm (e.g., 100 channels spaced by 50 GHz), it is approximately equal to 0.06 \cite{Poggiolini2012}. The parameter $\epsilon$ may also be computed from the approximation \cite[eq. (23)]{Poggiolini2012}.

\section{Optimization algorithms} \label{app:opt}

The particle swarm algorithm (PSO) randomly initializes $R$ particles $X = [L_{\text{EDF}}, \mathcal{P}_1, \ldots, \mathcal{P}_N]^T$. As the optimization progresses, the direction and velocity of the $i$th particle is influenced by its best known position and also by the best known position found by other particles in the swarm:

\begin{align}
v_i &\leftarrow wv_i + \mu_1 a_i(p_{i, best} - X_i) + \mu_2b_i(s_{best} - X_i) \tag{velocity} \\
X_i &\leftarrow X_i + v_i \tag{location}
\end{align}
where $w$ is an inertial constant chosen uniformly at random in the interval $[0.1, 1.1]$, $\mu_1 = \mu_2 = 1.49$ are the adaptation constants, $a_i, b_i \sim \mathcal{U}[0, 1]$ are uniformly distributed random variables, $p_{i, best}$ is the best position visited by the $i$th particle, and $s_{best}$ is the best position visited by the swarm.

To speed up convergence and avoid local minima, it is critical to initialize the particles $X = [L_{\text{EDF}}, P_1, \ldots, P_N]$ to within close range of the optimal solution. From the nature of the problem, we can limit the particles to a very narrow range. The EDF length is limited from 0 to 20 m. Since the amplifier gain will be relatively close to the span attenuation $A(\lambda) = e^{\alpha_{SMF}l}$, we can compute the maximum input power to the amplifier that will allow this gain for a given pump power $P_p$. This follows from conservation of energy \cite[eq. 5.3]{Desurvire1995}: 

\begin{equation}
P_n < \frac{1}{\bar{N}}\frac{\lambda_pP_p}{\lambda_nA(\lambda)},
\end{equation}
where $\lambda_p$ is the pump wavelength, $\lambda_n$ is the signal wavelength, and $\bar{N}$ is the expected number of WDM channels that will be transmitted. The minimum power is assumed to be 10 dB below this maximum value.

When nonlinear noise power is small, the solution found by the PSO does not change for different particle initializations. However, the solutions found by PSO when nonlinear noise is not negligible exhibit some small and undesired variability. To overcome this problem, after the PSO converges, we continue the optimization using the saddle-free Newton's method \cite{Dauphin2014}. According to this algorithm, the adaptation step $X \leftarrow X + \Delta X$ is given by
\begin{equation} \label{eq:sfn-step}
\Delta X = -\mu |H|^{-1}\nabla C,
\end{equation}
where $\mu$ is the adaptation constant, $\nabla C$ is the gradient of the capacity in \eqref{eq:capacity} with respect to $X$, and $H$ is the Hessian matrix, i.e., the matrix of second derivatives of $C$ with respect to $X$. The absolute value notation in \eqref{eq:sfn-step} means that $|H|$ is obtained by replacing the eigenvalues of $H$ with their absolute values. 

Both the gradient and the Hessian can be derived analytically by using the semi-analytical model given in equations \eqref{eq:semi_analytical_gain} and \eqref{eq:out-photon-flux}. However, we compute the gradient analytically and compute the Hessian numerically using finite differences of the gradient. 

\section*{Acknowledgments}
\noindent The authors are grateful for the valuable discussions with Marcio Freitas and Ian Roberts.

\bibliographystyle{ieeetr}
\bibliography{bib}

\end{document}